\documentstyle[aps]{revtex}
\def\jcite#1#2{#1\cite{#2}}
\def\jepsfbox#1{\typeout{#1} \epsfbox{#1}}

\def\plottwo#1#2{\centering \leavevmode
\epsfxsize=.45\columnwidth \jepsfbox{#1} \hfil
\epsfxsize=.45\columnwidth \jepsfbox{#2}}

\def\rmmat#1{{\hbox{\rm #1}}}
\def\rmscr#1{\rmmat{\scriptsize #1}}
\newcommand{\comment}[1]{}
\newcommand{\be}{\begin{equation}}
\newcommand{\ee}{\end{equation}}
\newcommand{\ba}{\begin{eqnarray}}
\newcommand{\ea}{\end{eqnarray}}
\newcommand{\bt}{\begin{table} \begin{center}}
\newcommand{\et}{\end{center} \end{table}}
\def\eqref#1{Eq.~\ref{eq:#1}}
\def\figref#1{Fig.~\ref{fig:#1}}
\def\tabref#1{Tab.~\ref{tab:#1}}
\input epsf
\def\d{{\rm d}}

\def\rarrow{\rightarrow}

\def\tilde{\mathaccent"365}			

\def\lta{\mathrel{\hbox{\rlap{\hbox{\lower4pt\hbox{$\sim$}}}\hbox{$<$}}}}
\def\gta{\mathrel{\hbox{\rlap{\hbox{\lower4pt\hbox{$\sim$}}}\hbox{$>$}}}}
\def\lsim{\mathrel{\hbox{\rlap{\hbox{\lower4pt\hbox{$\sim$}}}\hbox{$<$}}}}
\def\gsim{\mathrel{\hbox{\rlap{\hbox{\lower4pt\hbox{$\sim$}}}\hbox{$>$}}}}

\def\s{\ifmmode \widetilde \else \~\fi} 
\def\={\overline}
\def\etal{{\it et al.\ }}
\def\eg{{\it e.g.~}}
\def\ie{{\it i.e.~}}

\begin{document}
\title{Magnetically Catalyzed Fusion}
\author{Jeremy S. Heyl \and 
Lars Hernquist}
\address{Lick Observatory, 
University of California, Santa Cruz, California 95064, USA}
\maketitle \draft
\begin{abstract}
We calculate the reaction cross-sections for the fusion of hydrogen
and deuterium in strong magnetic fields as are believed to exist in
the atmospheres of neutron stars.  We find that in the presence of a
strong magnetic field ($B \gsim 10^{12}$G), the reaction rates are many
orders of magnitude higher than in the unmagnetized case.  The fusion
of both protons and deuterons are important over a
neutron star's lifetime for ultrastrong magnetic fields ($B \sim
10^{16}$G).  The enhancement may have dramatic effects on
thermonuclear runaways and bursts on the surfaces of neutron stars.
\end{abstract}

\section{Introduction: Atomic Structure in an Intense Magnetic Field}

In large magnetic fields a hydrogen atom is compressed both perpendicular
and parallel to the field direction.  In a sufficiently strong magnetic
field (\hbox{$B \gsim 10^{12}$~G}), the Schr\"odinger equation for
the dynamics of the electron separates into axial and perpendicular 
(azimuthal and radial) equations.  As the potential is axisymmetric
around the direction of the magnetic field, we expect no azimuthal
dependence in the ground-state wavefunction of the electron.

In the direction perpendicular to the magnetic field, the wavefunction
can be obtained exactly \cite{Land77}.  This azimuthal wavefunction
is denoted by two quantum numbers $n$ and $m$.  Here we take $n=0$, as
the $n > 0$ solutions are less bound and therefore provide less
shielding.

The perpendicular wavefunction has the same form as the Landau
wavefunction for an electron in a magnetic field:
\be
R_{0m} (\rho,\theta) = {1 \over \sqrt{2^{m+1} \pi m!} a_H^{m+1}} \rho^m \exp \left ( -
{\rho^2 \over 4 a_H^2} \right ) e^{i m \theta}
\ee
\be
R^2_{0m} (\rho) = {(-1)^m \over 2 \pi m! } { 1 \over a_H^2} 
\left . \left ( \d \over \d \beta \right )^m \left [
\exp \left ( - \beta {\rho^2 \over 2 a_H^2} \right ) \right ]
\right |_{\beta=1}
\ee
where 
\be
a_H = \sqrt{\hbar/m_e \omega_H} = \sqrt{\hbar c / |e| H}
\ee

\subsection{The axial wavefunction}

Along the direction of the magnetic field, the electron experiences an
effective potential,
\be
V_{\rmscr{eff},0m}(z) = \langle R|V(r)|R \rangle  = \int_0^\infty -{e^2 \over \sqrt{z^2+\rho^2}}
R^2_{0m}(\rho) 2 \pi \rho\,\d \rho.
\ee
Performing the integral yields
\be
V_{\rmscr{eff},0m}(z) = -{e^2 \over a_H} \sqrt{\pi/2} {(-1)^m\over m! } \left. \left ( \d \over \d \beta
\right )^m \left [ {1 \over \sqrt{\beta}} \exp(\beta z^2/2 a_H^2) 
\rmmat{erfc}(\sqrt{\beta} |z|/\sqrt{2} a_H) \right ] \right |_{\beta=1}
\ee
which for large $z$ approaches  $-e^2/z$.  The Schr\"odinger equation
with this potential is not analytically solvable.  We can note certain
features of the desired solution.  Because, $V_\rmscr{eff}$ is
everywhere finite, both the wavefunction and its first derivative must
be continuous.  Rather than solve the equation directly, we use a variational
principle, which constrains the ground-state wavefunction ($\nu=0$)
along the magnetic field for the given values of $n$ and $m$.  The
index $\nu$
counts the number of nodes in the axial wavefunction. As with the $n > 0$ 
states, the $\nu>0$ states are barely bound compared to the $\nu=0$ state.

Looking at the radial wavefunction, we take the wavefunction
along the z-axis to be a Gaussian as well:
\be
Z(z) = {1 \over \sqrt[4]{2 \pi} \sqrt{a_z}} \exp \left (-{z^2 \over 4 a_z^2}
\right )
\ee
We must minimize the integral,
\be
I = \langle Z|H_\rmscr{eff}|Z \rangle = \int_{-\infty}^\infty \left [ (\hbar^2/2 m_e) (\nabla_z Z)^2 + 
V_\rmscr{eff} Z^2 \right ] \d z
\ee
For this problem the integral is (using the definition of $Z$ and $m=0$),
\be
I =  2 \int_0^\infty Z^2 \left [ {\hbar^2 \over 2 m_e} {a_H^3 u^2 \over 4 a_z^4} - 
e^2 \sqrt{\pi/2} \exp(u^2/2) 
\rmmat{erfc}(u/\sqrt{2}) \right ] \d u
\ee
where we have substituted $u=z/a_H$.  Next, we use the definition of $Z$,
\be
I =  2 \left [ {\hbar^2 \over 2 m_e} {a_H^3 \over 4 \sqrt{2\pi} a_z^5} 
     \int_0^\infty u^2 \exp \left (-{u^2 a_H^2 \over 2 a_z^2} \right ) \d u -
{e^2 \over 2 a_z} \int_0^\infty \exp(u^2/2) 
     \exp \left (-{u^2 a_H^2 \over 2 a_z^2} \right ) \rmmat{erfc}(u/\sqrt{2}) 
     \d u \right ].
\ee
The first integral is tractable yielding the quantity to be minimized,
\be
I = 2 \left [ {\hbar^2 \over 16 m_e a_H^2} {1 \over \alpha^2} - {e^2 \over 2 a_H}
{1 \over \alpha} \int_0^\infty \exp\left( {u^2 \over 2} (1-1/\alpha^2) \right) 
\rmmat{erfc}(u/\sqrt{2}) \,\d u \right ],
\ee
with respect to $\alpha=a_z/a_H$.  This minimization yields a value
of $a_z$.  \tabref{gaures} lists the results for the
minimization for several magnetic field strengths and compares them with
the eigenvalues for the energy of the bound state derived by
\jcite{Ruder \etal}{Rude94}.
Ruder \etal use a series of basis functions to solve the Schr\"odinger
equation.  
\begin{table}
\caption{The results of the minimization.}
\label{tab:gaures}
\begin{tabular}{r|rr|rrrr}
 & \multicolumn{2}{c|}{Ruder \etal} & \multicolumn{4}{c}{Our Results} \\
\multicolumn{1}{c|}{$B$ (G)} & \multicolumn{1}{c}{$E_{m=0}$ (Ry)} &
\multicolumn{1}{c|}{$E_{m=1}$ (Ry)} &
\multicolumn{1}{c}{$\alpha_{m=0}$} & \multicolumn{1}{c}{$E_{m=0}$ (Ry)} & 
\multicolumn{1}{c}{$\alpha_{m=1}$} & \multicolumn{1}{c}{$E_{m=1}$ (Ry)} \\
\hline
$4.7 \times 10^9$ & 2.04 & 1.20 & 1.14 & 1.77 & 1.59 & 1.15 \\
$4.7 \times 10^{10}$ & 4.43 & 2.93 & 2.00 & 4.18 & 2.65 & 2.85 \\
$4.7 \times 10^{11}$ & 9.45 & 6.69 & 3.79 & 8.91 & 4.79 & 6.44 \\
$4.7 \times 10^{12}$ & 18.6 & 13.9 & 7.77 & 17.1 & 9.35 & 13.0 \\
$4.7 \times 10^{13}$ &  -~~ & -~~ & 17.3 & 29.6 & 20.0 & 23.6 \\
$4.7 \times 10^{14}$ &  -~~ & -~~ & 38.1 & 47.0 & 46.1 & 38.1 \\
$4.7 \times 10^{15}$ &  -~~ & -~~ & 102. & 69.6 & 113. & 59.1  \\
$4.7 \times 10^{16}$ &  -~~ & -~~ & 265. & 97.7 & 288. & 84.8 \\
\end{tabular}                            
\end{table}
Our binding energies fall short of theirs by approximately twenty
percent, because we are restricted by our trial wavefunction.  We also
tried a sum of Gaussians but this added degree of freedom did not
yield significantly more tightly bound wavefunctions.

Using the results of the minimization, the electron probability density is
\be
\rho(r,z) = {1 \over a_H^2 a_z (2\pi)^{3/2}} \exp \left [ -\left (
{r^2\over 2 a_H^2} + {z^2\over 2 a_z^2} \right ) \right ]
\label{eq:rhocloud}
\ee
where we have combined the two Gaussians in a revealing fashion.  The
quadrupole moment of the distribution is given by $Q=2
a_H^2(\alpha^2-1)$.  Next we define a quantity
\be
n^2 = r^2 + \left ( {a_H \over a_z} \right )^2 z^2 = r^2 + {z^2 \over \alpha^2}
\ee
and recast the previous equation into the form
\be
\rho(r,z) = {1 \over a_H^2 a_z (2\pi)^{3/2}} \exp \left (
-{n^2\over 2 a_H^2} \right ).
\ee

\subsection{The screening potential}

\begin{figure}
\plottwo{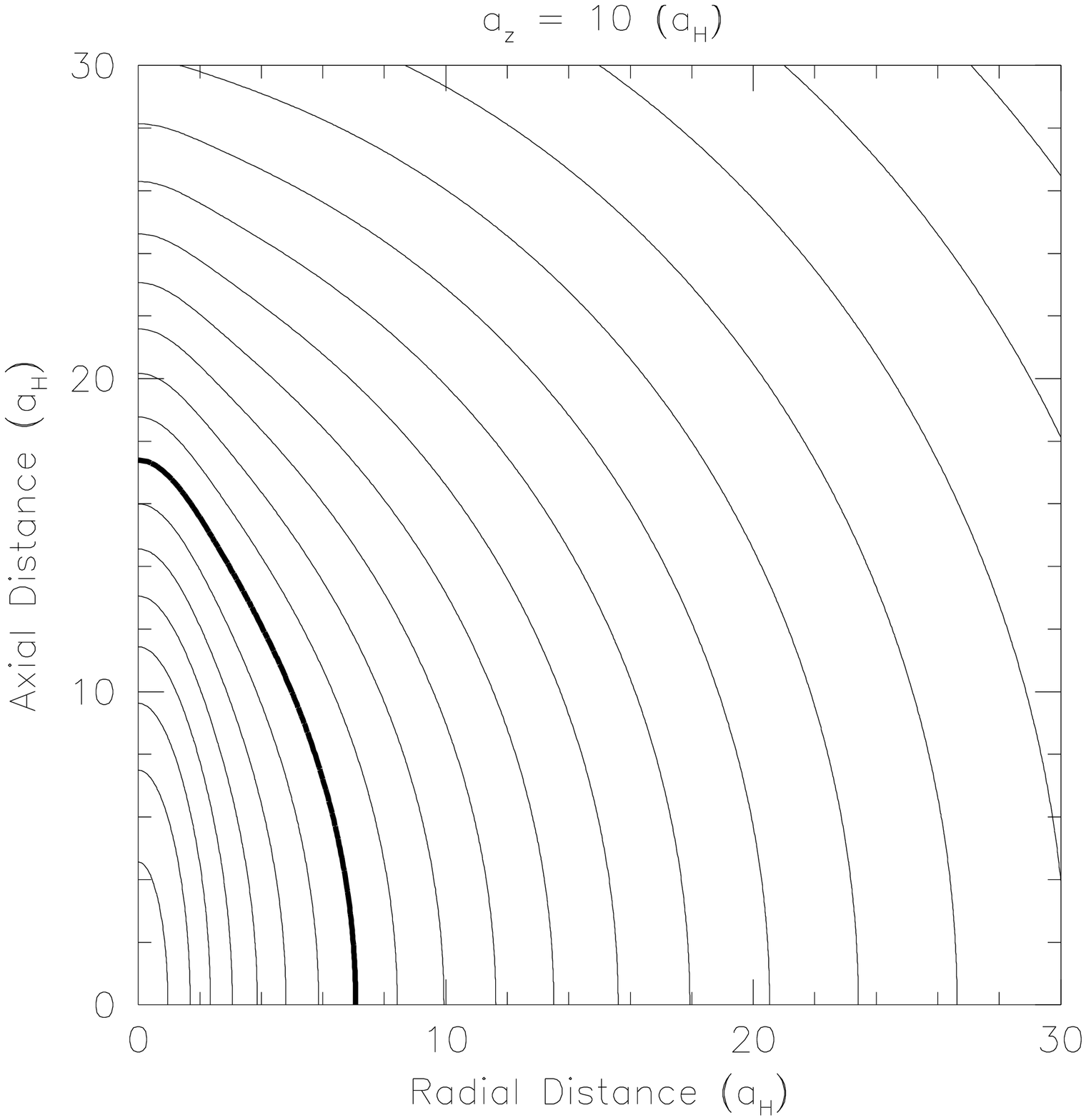}{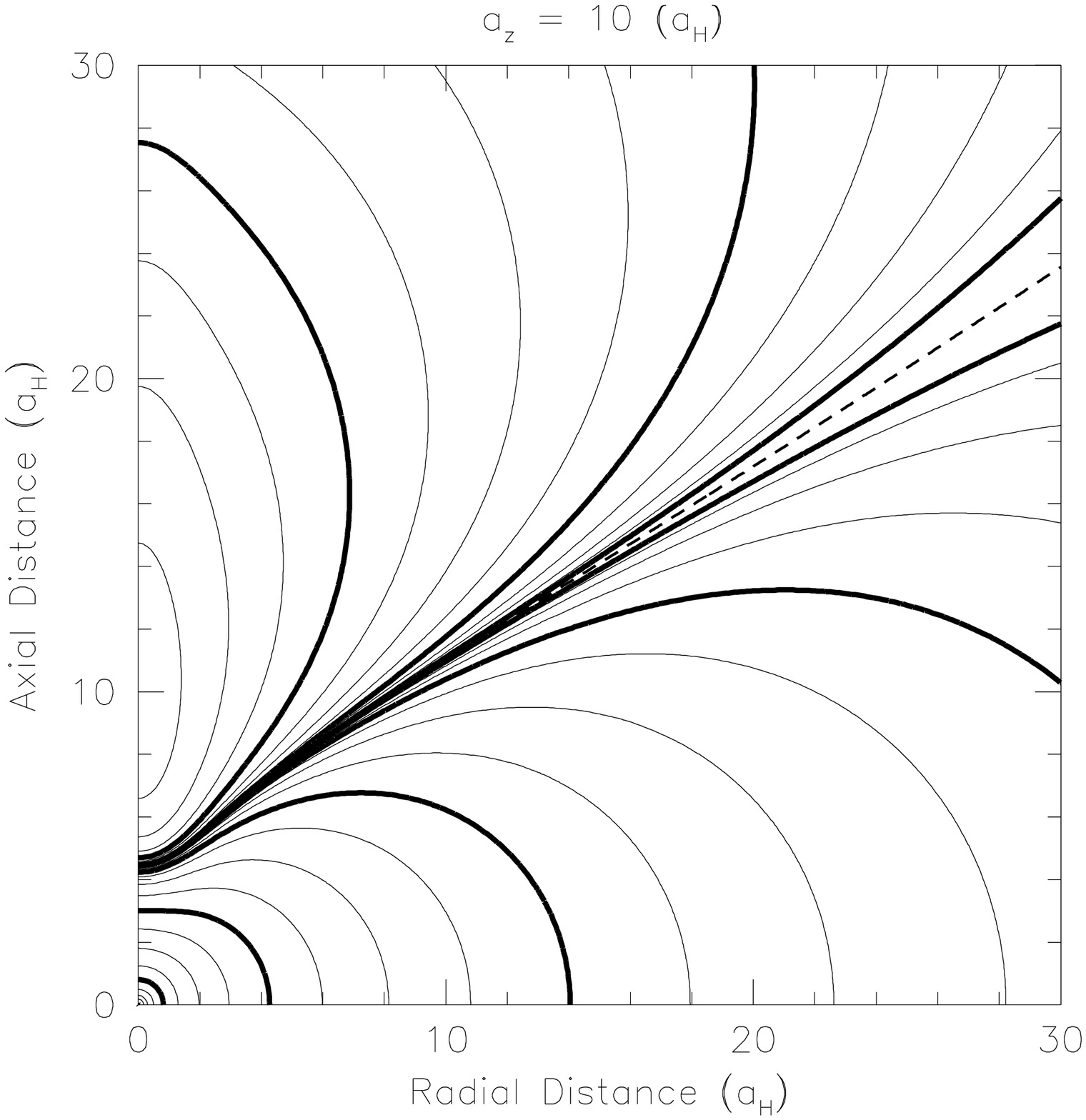}
\caption{The left panel depicts the screening potential as a function
of radius and z-position.  The right panel shows the total potential
experienced by an incoming proton.  The dashed contour denotes zero
potential.  The other contours are logarithmically spaced. In the left
panel the bold contour traces a potential of $0.1 e^2/a_H$.  The
contour levels increase toward the center with a spacing of $10^{1/20}$
In the right panel the bold contours trace potentials of $\pm 10^{-4},
10^{-3} \ldots e^2/a_H$.}
\label{fig:potrz}
\end{figure}

When solving gravitational problems one often looks for electrostatic 
analogues.  Here, we look for a gravitational analogue to an
electrostatic problem.  The density of the electron is constant on
concentric, similar homoemoids.  For this density distribution the
potential is directly solvable \cite{Binn87}
\be
\label{eq:phibin}
\Phi({\vec x}) = - \pi G \left ( {a_2 a_3 \over a_1} \right )
	\int_0^\infty {{\psi(\infty) - \psi(m)} \over 
	\sqrt{(\tau+a_1^2)(\tau+a_2^2)(\tau+a_3^2)}} \,\d\tau
\ee
where we have the following auxiliary definitions:
\be
m^2 = a_1^2 \sum_{i=1}^3 {x_i^2 \over a_i^2 + \tau}
\ee
and
\be
\psi(m) = \int_0^{m^2} \rho(m^2) \,\d m^2.
\ee
In our case, we use $G=-e^2$, $a_1=a_2=a_H$, $a_3=a_z$ and
\be
\psi(m) = {1 \over a_z \pi \sqrt{2 \pi}} \left [ 1 - \exp \left (-{m^2
\over 2 a_H^2} \right ) \right ].
\ee
Substituting these results into \eqref{phibin} yields
\be
\Phi({\vec x}) = {1 \over \sqrt{2 \pi}} e^2 
	\int_0^\infty {\exp \left [ -{1 \over 2} \left (
	{r^2 \over a_H^2+\tau} + {z^2 \over a_z^2 + \tau} \right )
	\right ] \over 
	(\tau+a_H^2)\sqrt{\tau+a_z^2}} \,\d\tau. 
\ee
We change variables to simplify the integral.  Using the natural units
of the problem, we let ${\bar r} = r/a_H$, ${\bar z} = z/a_H$, and 
$u = \tau/a_H^2$.  This gives the new equation
\be
\Phi({\vec x})={e^2 \over a_H} 	{1 \over \sqrt{2 \pi}} 
	\int_0^\infty {\exp \left [ -{1 \over 2} \left (
	{{\bar r}^2 \over 1+u} + {{\bar z}^2 \over \alpha^2 + u} \right )
	\right ] \over 
	(1+u)\sqrt{\alpha^2+u}} \,\d u
\ee
where we again use the previous definition of $\alpha$.  The potential
at the center of the electron cloud ($r=0, z=0$) is given by
\be
	\Phi(0,0) = {e^2 \over a_H} 	
	{2 \over \sqrt{2 \pi}} { \ln \left (\alpha +
	\sqrt{\alpha^2-1} \right ) \over \sqrt{\alpha^2-1}}.
\ee
Moving away from the origin a change of variables is
useful when evaluating the integral.  Let
\be
	v = {1 \over 1 + u}.
\ee
The integral becomes 
\be
\Phi({\vec x})={1 \over \sqrt{2 \pi}} {e^2 \over a_H} 	
	\int_0^1 {\exp \left [ -{1 \over 2} \left (
	{\bar r}^2 v + {\bar z}^2 { v \over 1 + (\alpha^2-1) v }
	\right ) \right ] \over \sqrt{(\alpha^2-1)v^2+v}}\,\d v.
\label{eq:potsimp}
\ee
As an example we present results for $B = 9.8 \times 10^{12}$ G.   For
this field
strength
$a_H \approx 10^{-12}$ m and $a_z \approx 10^{-11}$ m, so $\alpha =
10$.  The range of the nuclear force is approximately $10^{-15}$ m or $0.001
a_H$.  \figref{potrz} depicts the potential in units
of $e^2/a_H$ for this configuration.  The central potential is
approximately $0.25 e^2/a_H$ and drops quickly in the radial
direction.  In the axial direction, the potential forms a ``core'' 

The total potential of the electron cloud and the proton may be
approximated by the quadrupole formula
\be
V\left ({\vec x}\right ) \approx -\frac{\alpha^2-1}{2} \frac{e
a_H^2}{r^3} \left (3 \cos^2 \phi -1\right)
\label{eq:qpot}
\ee
for large separations.

\subsection{The Cloud-Cloud Potential}

When we consider the interaction between the two electron clouds
surrounding the protons, we must account not only for their electrical
potential but also the antisymmetry of the mutual electron
wavefunction.  Because of the strong ambient field, we expect that
both electron spins will be aligned with the field, so the spatial
component of the wavefunction must be antisymmetric.  That is,
\be
\Psi({\vec x}_1,{\vec x}_2) =
\frac{1}{\sqrt{2}} \left ( \psi_1({\vec x}_1) \psi_2({\vec x}_2) -
\psi_1({\vec x}_2) \psi_2({\vec x}_1) \right )
\ee
where 
\be
\psi_1({\vec x}) = \sqrt{\rho({\vec x})}
\rmmat{~and~} \psi_2({\vec x}) = \psi_1({\vec x} - {\vec x}_0)
\ee
with $\rho({\vec x})$ given by \eqref{rhocloud} and
${\vec x}_0$ is the position of the center of the second electron
cloud.

The potential energy of the two electrons is given by (\eg \cite{Land3})
\def\Potcc{\Phi_\rmscr{cc}}
\begin{eqnarray}
\Potcc({\vec x}_0) &=& \int \int \frac{e^2}{|{\vec x}_1-{\vec x}_2|}
|\Psi({\vec x}_1,{\vec x}_2)|^2 \d^3 x_1 \d^3 x_2 \\
	      &=& A({\vec x}_0) - J({\vec x}_0)
\end{eqnarray}
where $A({\vec x}_0)$ and $J({\vec x}_0)$ are given by
\begin{eqnarray}
A({\vec x}_0) &=& \int \int \frac{e^2}{|{\vec x}_1-{\vec x}_2|}
\rho_1({\vec x_1}) \rho_2({\vec x_2}) \d^3 x_1 \d^3 x_2 \\
J({\vec x}_0) &=& \int \int \d^3 x_1 \d^3 x_2 
 \Biggr [ \frac{e^2}{|{\vec x}_1-{\vec x}_2|}
\rho_1({\vec x_1}) \rho_2({\vec x_2}) \nonumber \\*
& & \times \exp \left ( -\frac{1}{2} \left ( \frac{(x_1-x_2)x_0 +
(y_1-y_2)y_0}{a_H^2} + \frac{(z_1-z_2)z_0}{a_z^2} \right )\right )
\Biggr ] \\
&\approx& \int \int \d^3 x_1 \d^3 x_2 
 \Biggr ( \frac{e^2}{|{\vec x}_1-{\vec x}_2|}
\rho_1({\vec x_1}) \rho_2({\vec x_2}) \nonumber \\*
& & \times \exp \left [ -\frac{1}{2} \left ( \frac{x_0^2 +
y_0^2}{a_H^2} + \frac{z_0^2}{a_z^2} \right )\right ]
\Biggr ) \\
&\approx& A({\vec x}_0) \exp \left [ -\left (
{r_0^2\over 2 a_H^2} + {z_0^2\over 2 a_z^2} \right ) \right ]
\end{eqnarray}
where we have used the Gaussian form of $\rho({\vec x})$ to simplify
the expression for $J({\vec x}_0)$, and to obtain its approximate
value we replace $x_1-x_2$ by $x_0$ and similarly for the other
coordinates.  

To calculate the direct term of cloud-cloud potential ($A({\vec
x}_0$)) we will
take advantage of the
special form of the density distribution given in
\eqref{rhocloud}.  The direct term 
is in general
given by
\be
A\left ({\vec x}_0\right ) = \int \d^3 x_1 \rho\left ({\vec x}_1 - {\vec x}_0\right ) \Phi\left ({\vec x}_1\right )
\ee
which is simply the convolution of the density distribution with the
potential.  If we perform the Fourier transform of the right-hand side
we get
\be
A\left ({\vec x}_0\right ) = \int \d^3 k {\tilde\rho}\left ({\vec k}\right ) {\tilde\Phi}\left ({\vec k}\right )
e^{-i {\vec k}\cdot{\vec x}_0}.
\label{eq:potcc1}
\ee
Expressing the Poisson equation in Fourier space gives
\be
\Phi\left ({\vec x}\right ) = - 4 \pi \int \frac{\d^3 k}{(2\pi)^{3/2}}
\frac{{\tilde\rho}\left ({\vec k}\right )}{k^2} e^{-i {\vec k}\cdot{\vec
x}}.
\label{eq:potft}
\ee
Because the magnetic field induces the deformation of both electron
clouds, the clouds are aligned, and they have the same Fourier transforms; therefore,
\be
A\left ({\vec x}_0\right ) = -4 \pi \int \d^3 k
\frac{\left[{\tilde\rho}\left ({\vec k}\right )\right]^2}{k^2}
e^{-i {\vec k}\cdot{\vec x}_0}.
\label{eq:potcc2}
\ee
Because $\rho$ is a three-dimensional Gaussian, so is its Fourier
transform; consequently,
\be
\left[{\tilde\rho}\left ({\vec k}\right )\right]^2 = \frac{1}{(2\pi)^{3/2}}
{\tilde\rho}\left (\sqrt{2} {\vec k}\right )
\label{eq:rho2}
\ee
Combining \eqref{potcc2} and \eqref{rho2}, yields
\be
A\left ({\vec x}_0\right ) = -4 \pi \frac{1}{(2\pi)^{3/2}} \int \d^3 k
\frac{{\tilde\rho}\left (\sqrt{2} {\vec k}\right )}{k^2}
e^{-i {\vec k}\cdot{\vec x}_0}.
\label{eq:potcc3}
\ee
Performing a change of variables ${\vec l} = \sqrt{2} {\vec k}$ gives
\be
A\left ({\vec x}_0\right ) = -4 \pi \frac{1}{\sqrt{2} (2\pi)^{3/2}} \int \d^3 l
\frac{{\tilde\rho}\left ({\vec l}\right )}{l^2}
e^{-i {\vec l}\cdot{\vec x}_0/\sqrt{2}}.
\label{eq:potcc4}
\ee
Comparing this equation with \eqref{potft}, we get
\be
A\left ({\vec x}_0\right ) = \frac{1}{\sqrt{2}} \Phi\left (\frac{{\vec x}_0}{\sqrt{2}} \right).
\ee
Therefore, the total potential energy between two hydrogen atoms
separated by ${\vec x}$ in the
magnetic field is given by
\be
V\left ({\vec x}\right ) \approx \frac{e^2}{r}
+ \frac{1}{\sqrt{2}}\Phi\left (\frac{\vec x}{\sqrt{2}} \right)
\left (1 - \exp \left [ -\left (
{r^2\over 2 a_H^2} + {z^2\over 2 a_z^2} \right ) \right ]
\right) - 2 \Phi\left ({\vec x} \right)
\label{eq:potccfinal}
\ee
where $\Phi\left ({\vec x} \right)$ is simply the potential induced by
the Gaussian cloud of charge (\eqref{potsimp}).

Far from the atoms ($r\gg \alpha a_H$), the interaction energy may be
approximated by the quadrupole-quadrupole energy,
\be
V\left ({\vec x}\right ) \approx \frac{3}{4} \left(\alpha^2-1\right)^2 \frac{e^2 a_H^4}{r^5}
\left( 35 \cos^4\phi - 30 \cos^2\phi + 3 \right).
\label{eq:qqpot}
\ee
where $\phi$ is the angle relative to the symmetry axis of the atom.

\begin{figure}
\plottwo{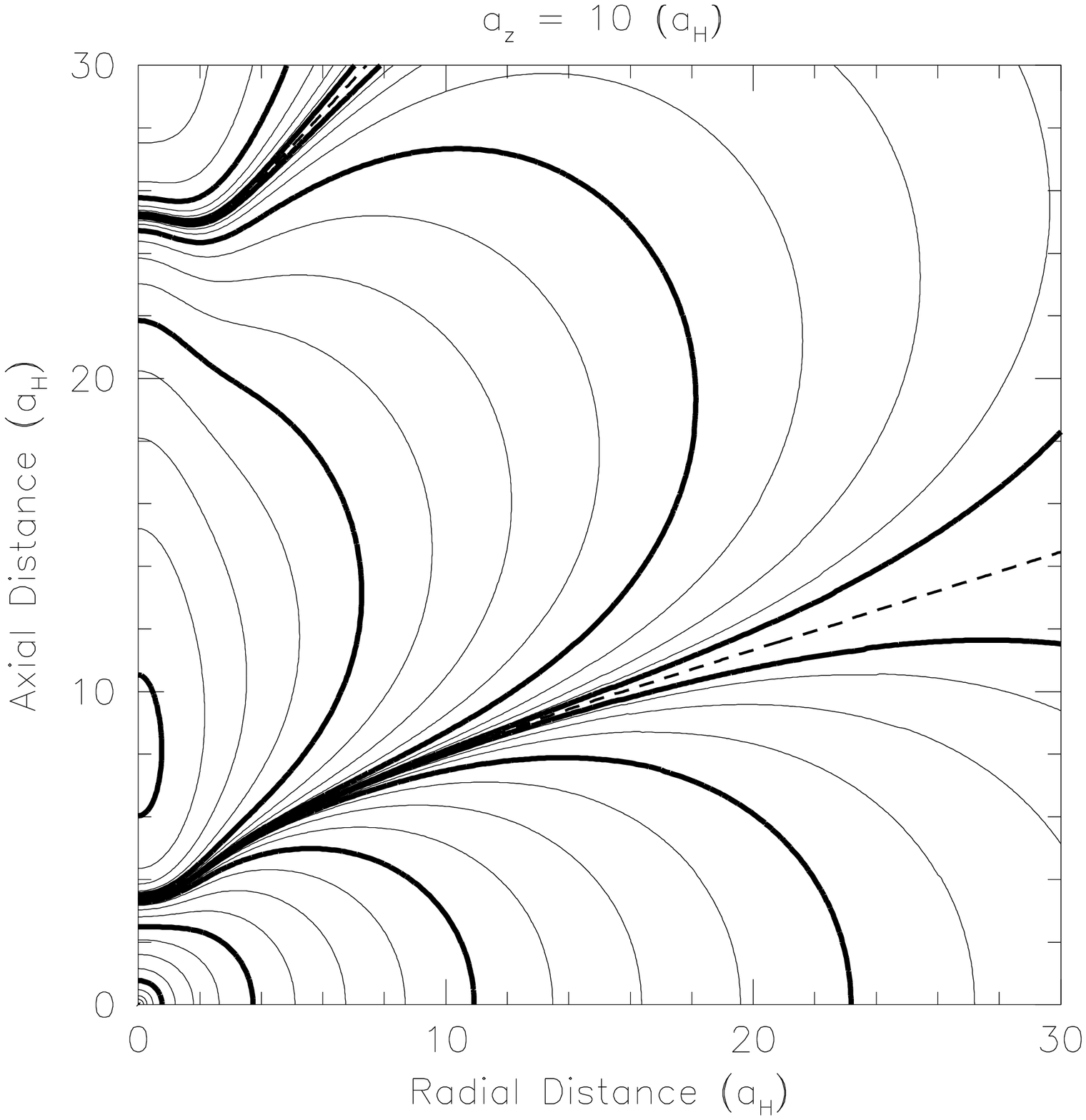}{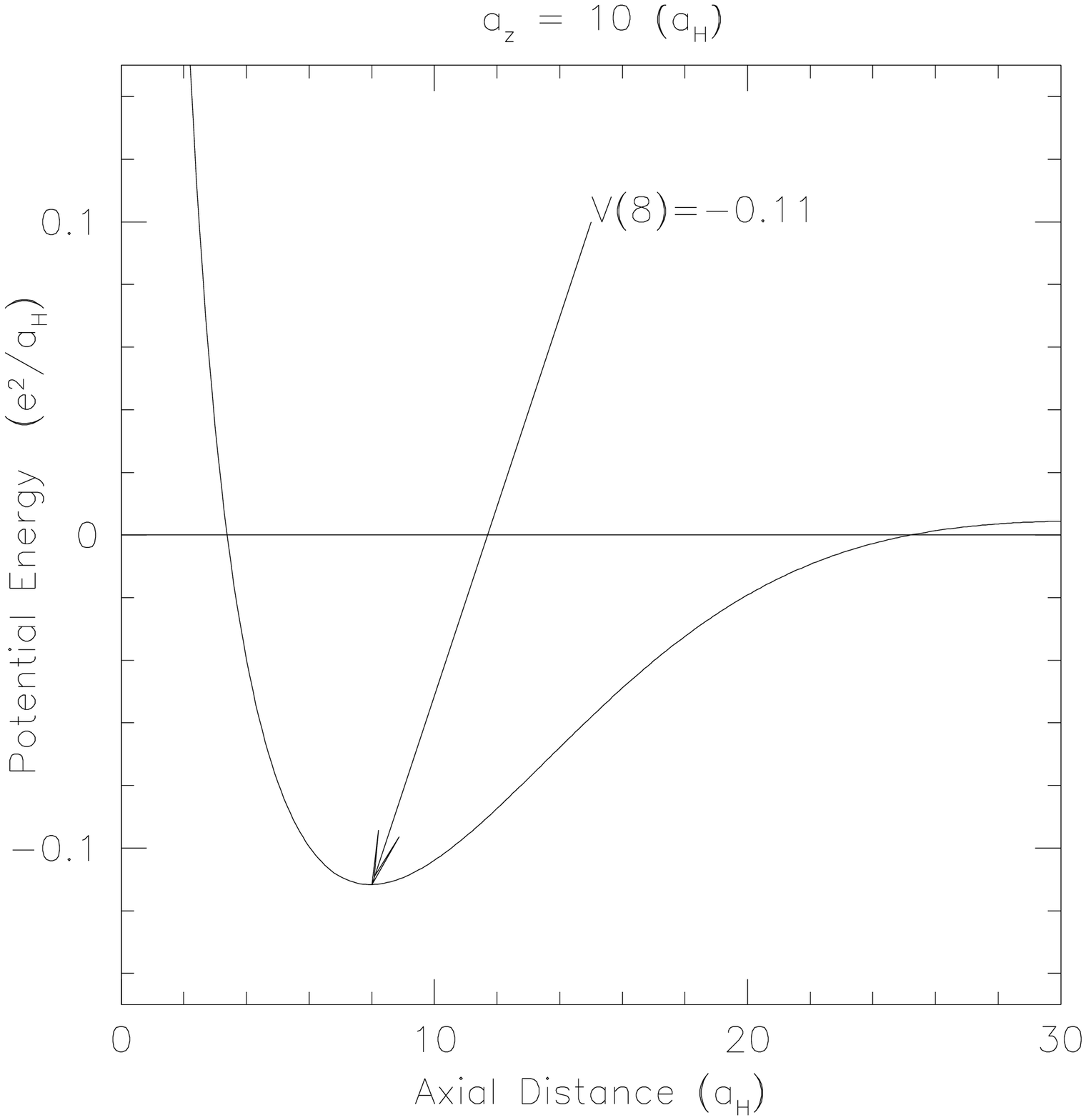} 
\caption{The figures depict the total potential energy between two
magnetized hydrogen atoms excluding the antisymmetrization energy.
For the left panel, the contour spacing is the same as in
right panel of \figref{potrz}.
The right panel illustrates the potential
along the axis of the magnetic field. }
\label{fig:directpot}
\end{figure}

\begin{figure}
\plottwo{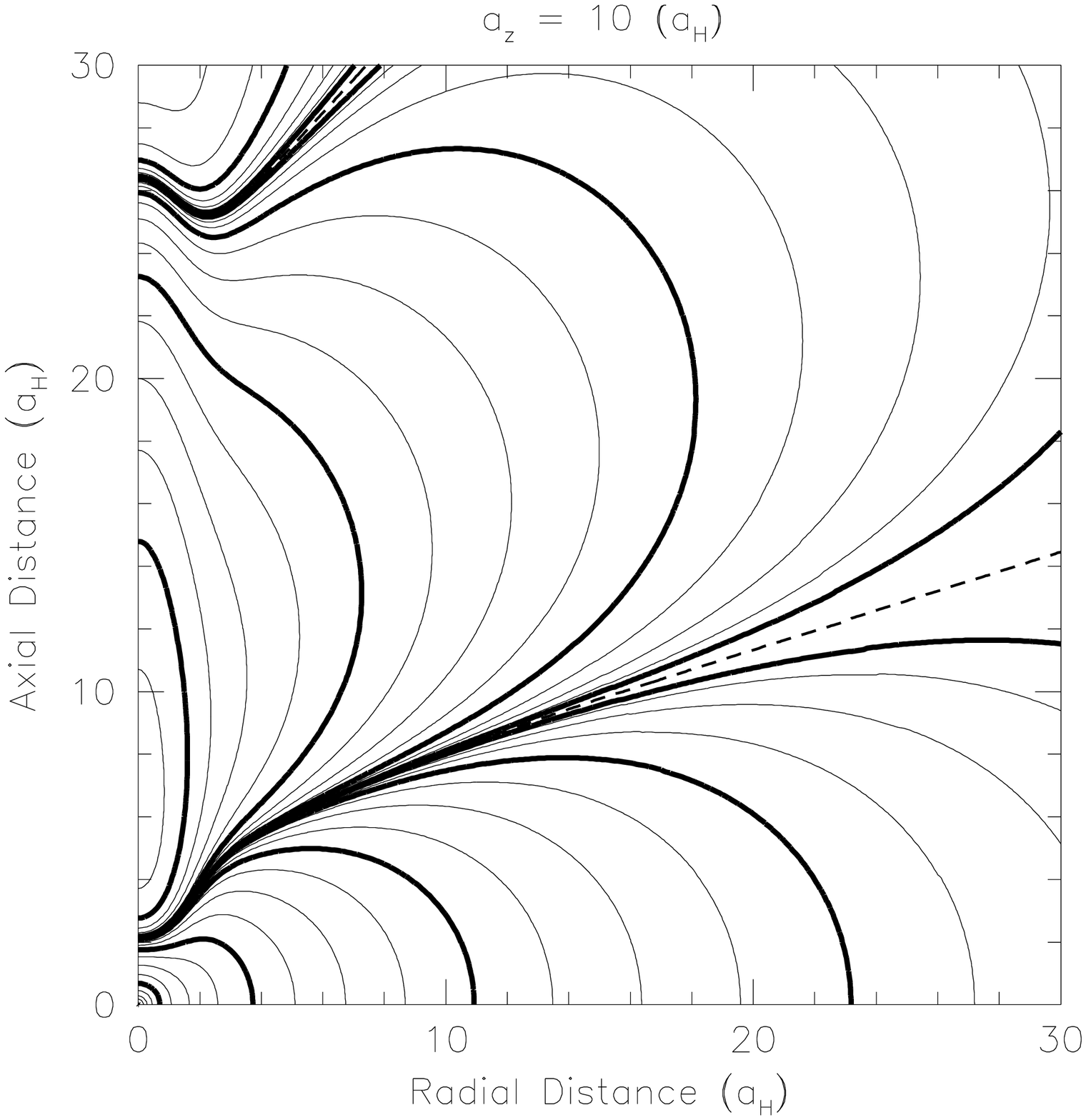}{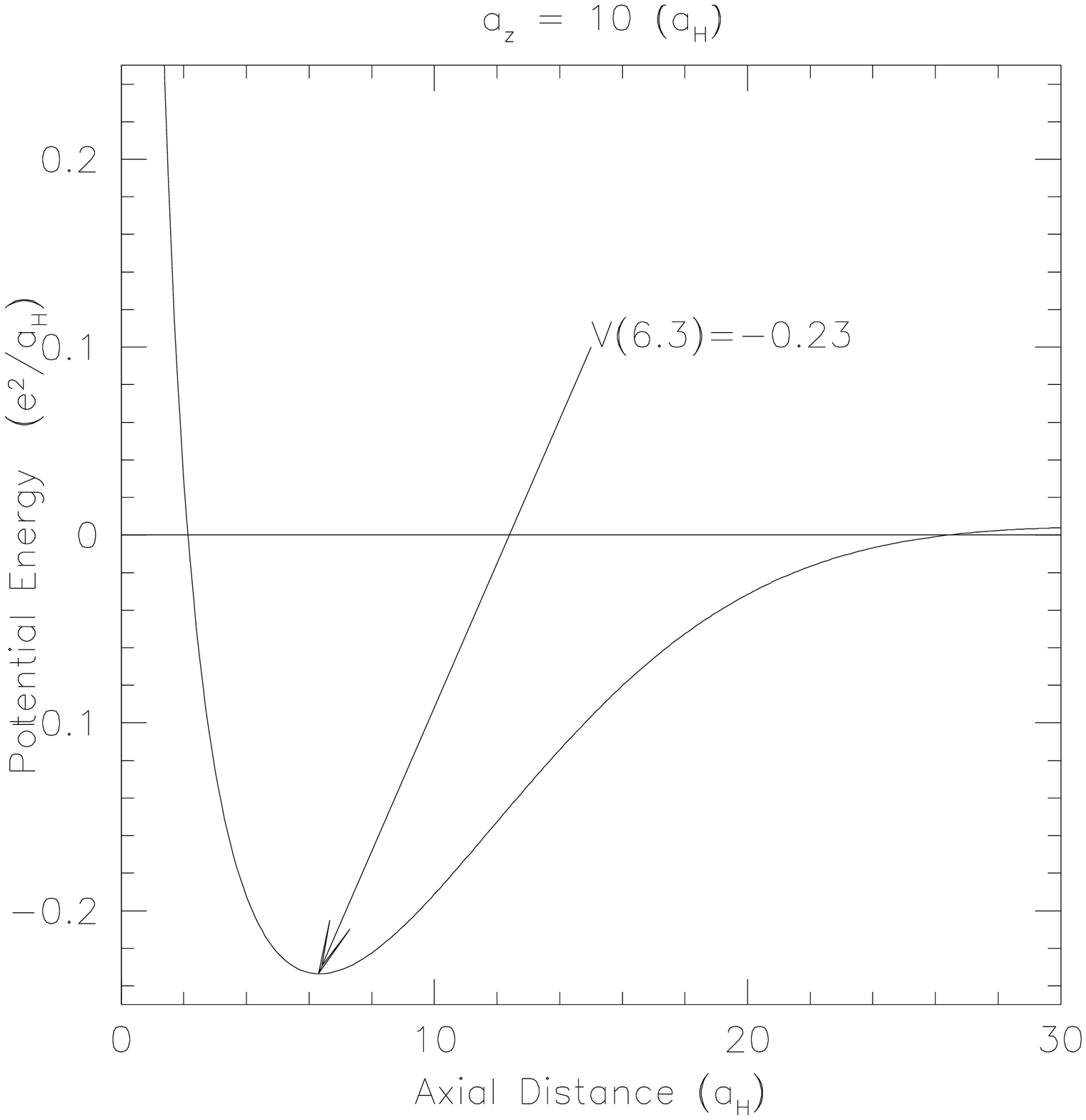}
\caption{The figures depict the total potential energy between two
magnetized hydrogen atoms including the antisymmetrization energy.
For the left panel, the contour spacing is the same as in
right panel of \figref{potrz}.
The right panel illustrates the potential
along the axis of the magnetic field. }
\label{fig:totalpot}
\end{figure}

\figref{directpot} and \figref{totalpot} depict
the total potential energy between
two hydrogen atoms in a magnetic field for the same magnetic-field
strength as \figref{potrz} ($B=9.8 \times 10^{12}$G).  A
comparison of the two figures illustrates that the
exchange term provides a slight attractive force between the two
electron clouds, because of the anticorrelation of the clouds.
At large separations, both potentials are well
approximated by the quadrupole-quadrupole formula
(\eqref{qqpot}).

\section{Estimating Reaction Rates}

In a fluid state, there will be three possible reaction channels,
\begin{itemize}
\item proton-proton dominates in hot, totally ionized gas
\item proton-atom dominates in nearly completely ionized gas
\item atom-atom dominates in neutral and partially ionized gas
\end{itemize}
For the first channel, we can use the standard thermonuclear reaction
rates (\eg \cite{Clay83}).  For the latter two channels we must
include the screening potentials that we calculated in the previous
section to determine the potential wall through which the interacting
particles must penetrate.

\subsection{The transmission probability}
In the WKB approximation, the probability to traverse through a potential wall
is
\begin{eqnarray}
|T|^2 & \approx & \exp \left ( -2 \int_\rmscr{Wall} \d r \sqrt{{2 m \over
\hbar^2} \left ( V(r) - E \right ) } \right ) \\
 & \approx & \exp \left ( -2 \sqrt{2 m a_H} {e \over \hbar} 
\int_\rmscr{Wall} \d u \sqrt{ V(r) - E \over e^2/a_H} \right ) \\
 & \approx & \exp \left ( -26.69 B_{12}^{-1/4} \int_\rmscr{Wall} \d u 
  \sqrt{ {\cal V}(u) - {\cal E}} \right )
\end{eqnarray}
where $B_{12}$ is the magnetic-field strength in units of $10^{12}$ G,
$u$ is the dimensionless radius $r/a_H$ and $\cal{E}$ and $\cal{V}$
are the dimensionless energy $E a_H/e^2$ and potential.

For the proton-atom channel, the
potential includes both that of the nucleus ${\cal V}=1/u$ and the
surrounding electron cloud (\eqref{potsimp}).  At large
distances from the nucleus, $u>>\alpha$, the total potential is well
approximated by the quadrupole (\eqref{qpot}).  For the
atom-atom channel, the total potential includes contributions from the
proton-proton, proton-electron  and electron-electron potentials
(\eqref{potccfinal}),
which is well approximated by the quadrupole-quadrupole formula
(\eqref{qqpot}) for large separations.

To calculate the transition probability, we use these quadrupole formulae to
approximate the potential for $u>4 \alpha$ and 
for $u<1/2$, we approximate the potential energy between the electron
clouds and the electron clouds and the protons by their central values.
This both speeds the calculation and reduces the numerical error.

\begin{figure}
\plottwo{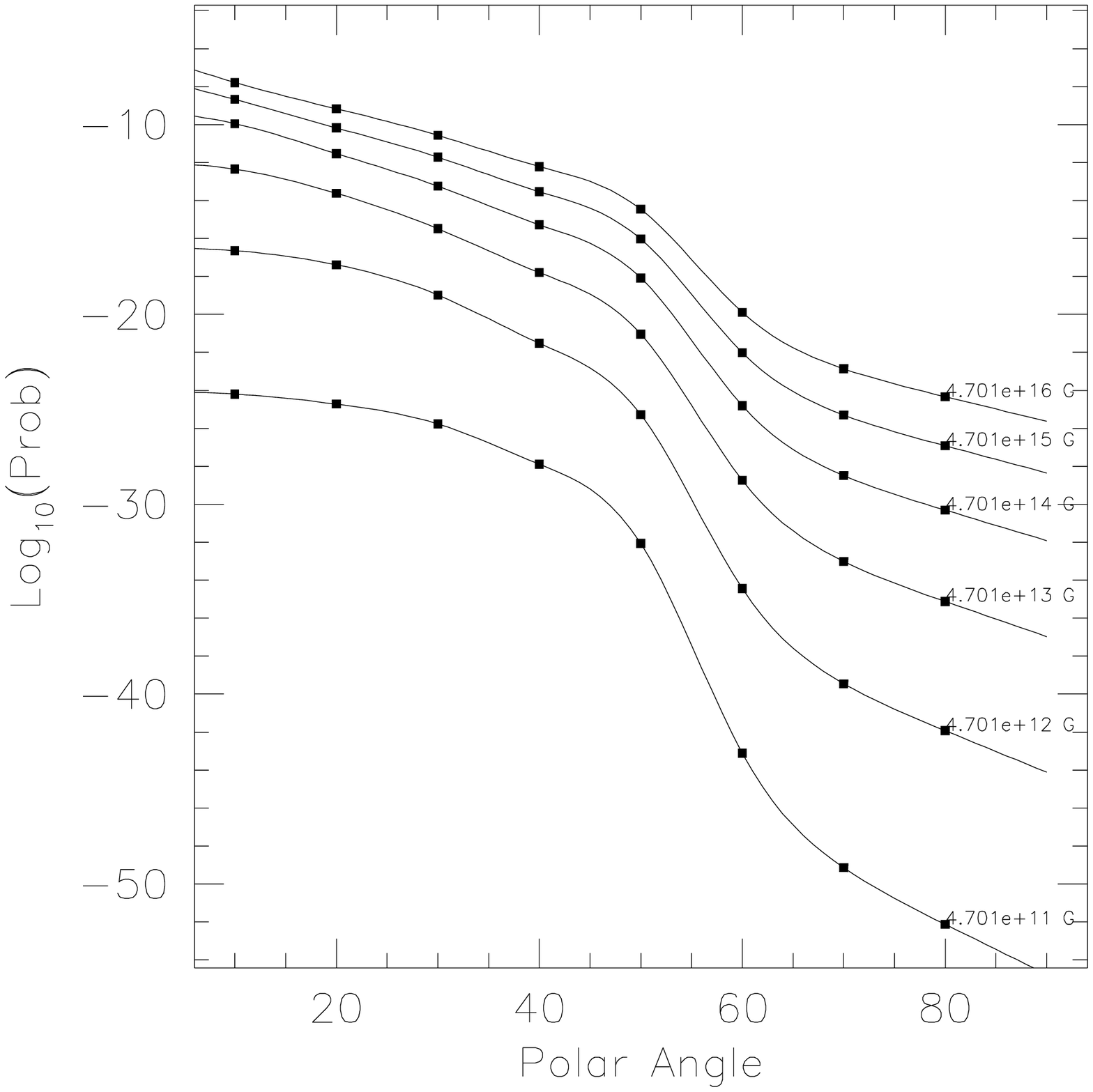}{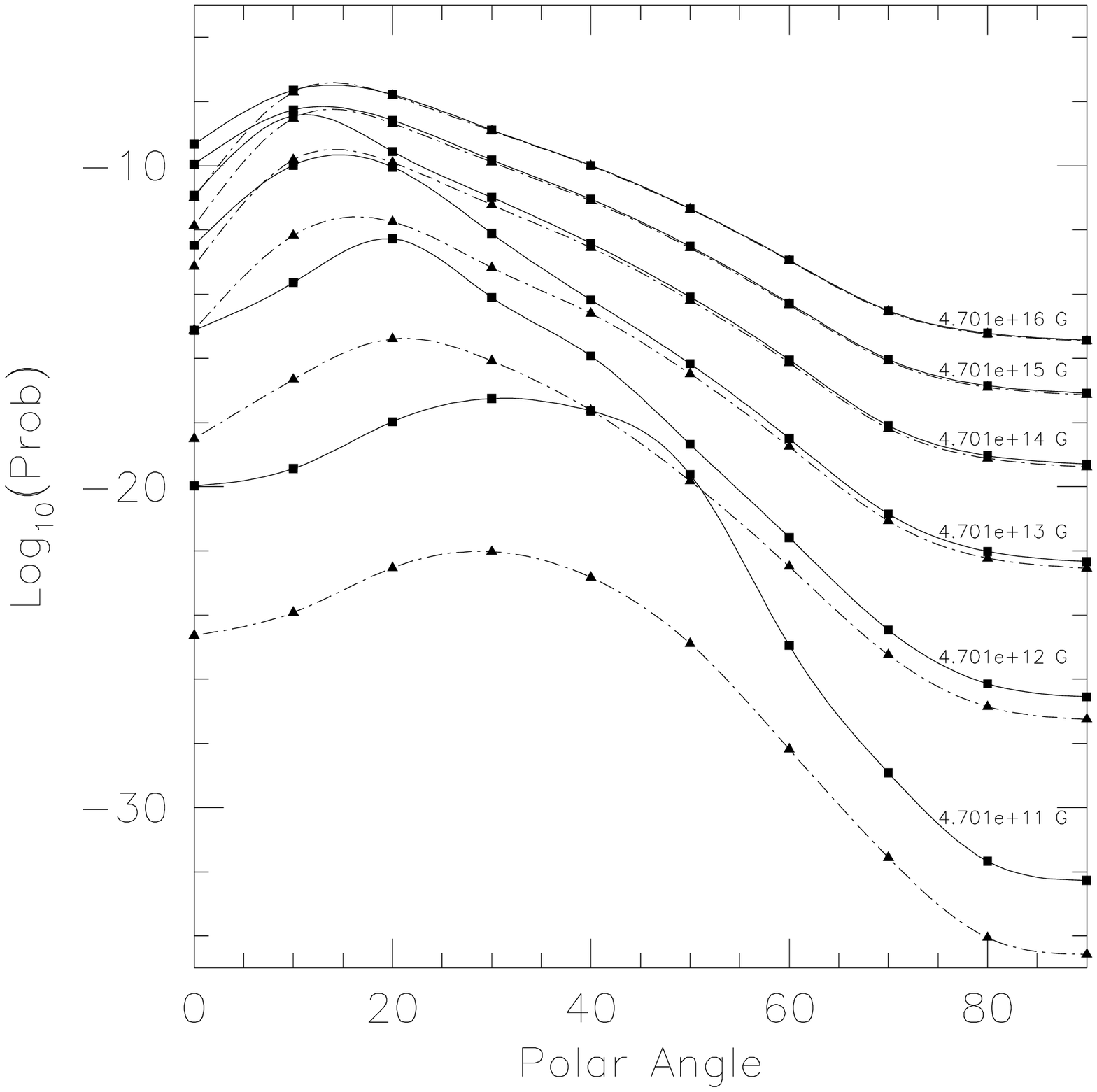}
\caption{The left panel depicts 
the transmission probability as a function of angle and magnetic 
field for a proton and an atom to interact at zero relative energy.
The right panel depicts the same probability for two atoms.  The solid
lines trace the probability if the antisymmetrization energy of the
electrons is considered.  The dashed lines show the probability
without antisymmetrization.}
\label{fig:transprob}
\end{figure}

\figref{transprob} traces the transmission probability for
protons to interact with atoms and atoms to interact with atoms at
zero relative energy as a function of angle and magnetic field.  In
the atom-proton case, the protons can most easily penetrate through
the mutual potential barrier along the axis of the magnetic field and
the penetration probability increases mark\'edly with the strength of
the magnetic field.  In the atom-atom case, we see that the maximum
transmission probability occurs at an angle to the field direction and
that with antisymmetrization of the electron density the transmission
probability increases dramatically.  For the reaction rate estimates
that follow we will account for the antisymmetrization energy of the
two electron clouds.

To translate this transmission probability into a cross section, we must 
average $|T|^2$ over a sphere and include the appropriate S-factor for the 
reaction where $S(E)$ is defined as
\be
S(E) = \sigma E |T|^2 \approx S_0 ( 1 + S_1 E)
\ee

In this way, the strong energy dependence of the reaction cross
section is removed.  For the reaction $^1$H$(p,e^+\nu)$D, $S_0 = 4.38
\times 10^{-25}$ MeV-barn and $S_1 = 11.2~\rmmat{MeV}^{-1}$
at low energies \cite{Bahc82}.  The reaction of the less abundant
deuterons with protons has a much larger S-factor of $S_0 = 2.5 \times
10^{-7}$ MeV-barn and $S_1 = 27.8~\rmmat{MeV}^{-1}$ \cite{Clay83}.

Given these definitions, the reaction proceeds at a rate of
\begin{eqnarray}
r_{12} & = & (1+\delta_{12})^{-1} n_1 n_2 \left < \sigma v \right > \\*
       & = &
        (1+\delta_{12})^{-1} n_1 n_2 \left ( {8 \over \mu \pi} \right )^{1/2} 
        S_0 (1 + S_1 k T) (k T)^{-1/2} |T|^2
\end{eqnarray}
where $\mu$ is the reduced mass of the reactants, and $n_1, n_2$ are their 
number densities.  $r_{12}$ has the units of reactions per unit time per unit
volume, so we can define a typical timescale for a reactant to be consumed
\be
\tau_{12} = n_1 / r_{12}.
\ee
We will use this timescale to assess the effectiveness of the screening in 
catalyzing the nuclear fusion reactions.  We also account for the increasing
excitation of the gas as the temperature increases and the onset of 
thermonuclear reactions above several million degrees.  

\subsection{The ground-state fraction}

The screening is much less effective if the electron is in an excited state, 
so we estimate the fraction of atoms in the ground state by first
calculating the ionization equilibrium and then the fraction of
neutral atoms in the ground state.

\jcite{Lai and Salpeter}{Lai95a} give the form of the Saha equation for hydrogen atoms,
electrons and protons in equilibrium in the presence of a quantizing
magnetic field.  Throughout this formalism, we use the natural units
of the problem, \ie $T$ is the temperature in units of $3.15 \times
10^5$ K, $M$ is the mass of the system in units of the electron's mass
(1840 for hydrogen and 3670 for deuterium), $b$ is the strength of the
magnetic field in units of $2.35 \times 10^9$ G and $n_g$ is the
number density of the gas in units of $6.76 \times 10^{24}
\rmmat{cm}^{-3}$.

We first look at the unexcited hydrogen atom.  For the partition
function of the ground state in a quantizing magnetic field,
\jcite{Lai and Salpeter}{Lai95a} give
\be
Z_\rmscr{ground}(H) \simeq n_g^{-1/3} \left({MT\over 2\pi}\right)^{1/2}
\exp\left({|E(H)|\over T}\right) Z_\perp
\ee
where  $E(H) = -0.16
l^2$ (the ground-state energy of the atom), $l=\ln b$, and
\begin{eqnarray}
Z_\perp & = & {n_g^{-2/3}\over (2\pi)^2}\int_0^{K_{\perp \rmscr{max}}}\!\!
2\pi K_{\perp}\d K_{\perp}\exp\left[-{E_\perp(K_\perp)\over T}\right]  \\
Z_{\perp} & \simeq & {n_g^{-2/3}\over 2\pi}\int_0^{K_{\perp \rmscr{max}}}\!\!
	K_{\perp}\d K_{\perp}\exp\left[-{\tau\over 2M_\perp T}
	\ln\left(1+{K_{\perp}^2\over\tau}\right)\right] \\
 & = & n_g^{-2/3} {M_\perp''T\over 2\pi},
\end{eqnarray}
where
$M_\perp=M+\xi b/l$ (with $\xi\simeq 2.8$) and
\be
\tau \simeq 0.64\, l \xi b 
\left[1+{M l \over \xi b} \right ]^2.
\ee

Here we have explicitly integrated to $K_{\perp \rmscr{max}}$, so we replace
$M_\perp'$ of \jcite{Lai and Salpeter}{Lai95a} with $M_\perp''$,
\be
M_\perp''=M_\perp' \left[ 1 - \left ( 1 + {K_{\perp \rmscr{max}}^2 \over \tau}
\right )^{-\tau/2 M_\perp' T} \right ]
\ee
and $M_\perp'$ is as given by \jcite{Lai and Salpeter}{Lai95a},
\be
M_\perp'=M_\perp\left(1-{2M_\perp T\over\tau}\right)^{-1}.
\ee
As $K_{\perp \rmscr{max}} \rarrow \infty$, $M_\perp'' \rarrow M_\perp'$ and we
recover the \jcite{Lai and Salpeter}{Lai95a} result.  $K_{\perp \rmscr{max}}$ is the upper limit
on the perpendicular momentum for the given state.  The electron
clouds of neighboring atoms should not
overlap; otherwise, the gas would become pressure ionized.  Therefore,
we take the size of the state, $R_K = K_\perp / b < R_g = n_g^{-1/3}$
as the defining condition on $K_{\perp \rmscr{max}}$.  We obtain
\be
K_{\perp \rmscr{max}}= b n_g^{-1/3}.
\ee

The total partition function of the neutral atom is given by
\be
Z(H) = Z_\rmscr{ground}(H) z_\nu(H) z_m(H)
\ee
where $z_\nu$ and $z_m$ are the partition functions for excitations of
the $\nu$ and $m$ quantum numbers respectively.  \jcite{Lai and Salpeter}{Lai95a} argue
that the $z_\nu(H) \simeq 1$ as these states are hardly occupied
relative to the ionized, $m>0$ and ground states.  For the contribution of the
$m>0$ states to the partition function, they obtain
\be
z_m(H)\simeq \left(1+e^{-b/MT}\right)
\sum_{m=0}^{\infty}{M_{\perp m}''\over M_\perp''}
\exp\left[-{1\over T}\left(0.16\,l^2-0.16\,l_m^2+m{b\over M}\right)\right],
\ee
where we have several additional auxiliary definitions:
\be
l_m=\ln \left ( {b \over 2m + 1} \right ),
\ee
and as with ground state we correct for $K_{\perp \rmscr{max}} <
\infty$ with
\be
M_{\perp m}''=M_{\perp m}' \left[ 1 - \left ( 1 + {K_{\perp \rmscr{max}}^2 \over \tau}
\right )^{-\tau_m/2 M_{\perp m}' T} \right ]
\ee
and $M_{\perp m}'$ is as given by \jcite{Lai and Salpeter}{Lai95a},
\be
M_{\perp m}'=M_{\perp m} \left(1-{2M_{\perp m}T\over
\tau_m}\right)^{-1}.
\ee
$M_{\perp m}$ is given by the following relation
\be
1-{M\over M_{\perp m}}\simeq {b\over M}\left[{m+1\over b/M+0.16\,l_m^2
-0.16\,l_{m+1}^2}-{m\over b/M+0.16\,l_{m-1}^2-0.16\,l_m^2}\right].
\ee
and we use the following additional definition
\be
\tau_m\simeq 0.64\, l_m (M_{\perp m} - M)
\left[1+{M \over M_{\perp m} - M} \right ]^2.
\ee

The ratio of the number of atoms in the ground state to the number of
neutral atoms is given by
\be
{ X_\rmscr{ground}(H) \over X(H)} = { Z_\rmscr{ground}(H) \over Z(H)}
= {1 \over z_m(H)}.
\label{eq:groundeq}
\ee

Next we calculate the ionization-recombination equilibrium.  
\begin{eqnarray}
{X(H)\over X_pX_e} & = & {Z(H)\over Z(p) Z(e)} \\
		   & \simeq & n_g\left({b\over 2\pi}\right)^{-2}\,
M_\perp'' \left({T\over 2\pi}\right)^{1/2} \tanh\left({b\over
2MT}\right) \nonumber \\
& & ~~\times \exp\left({|E(H)|\over
T}\right)z_m(H),
\label{eq:ioneq}
\end{eqnarray}
where $X(H)=n(H)/n_g$, $X_p=n_p/n_g$, $X_e=n_e/n_g$ are the number 
density fractions of the different species.  

\begin{figure}
\plottwo{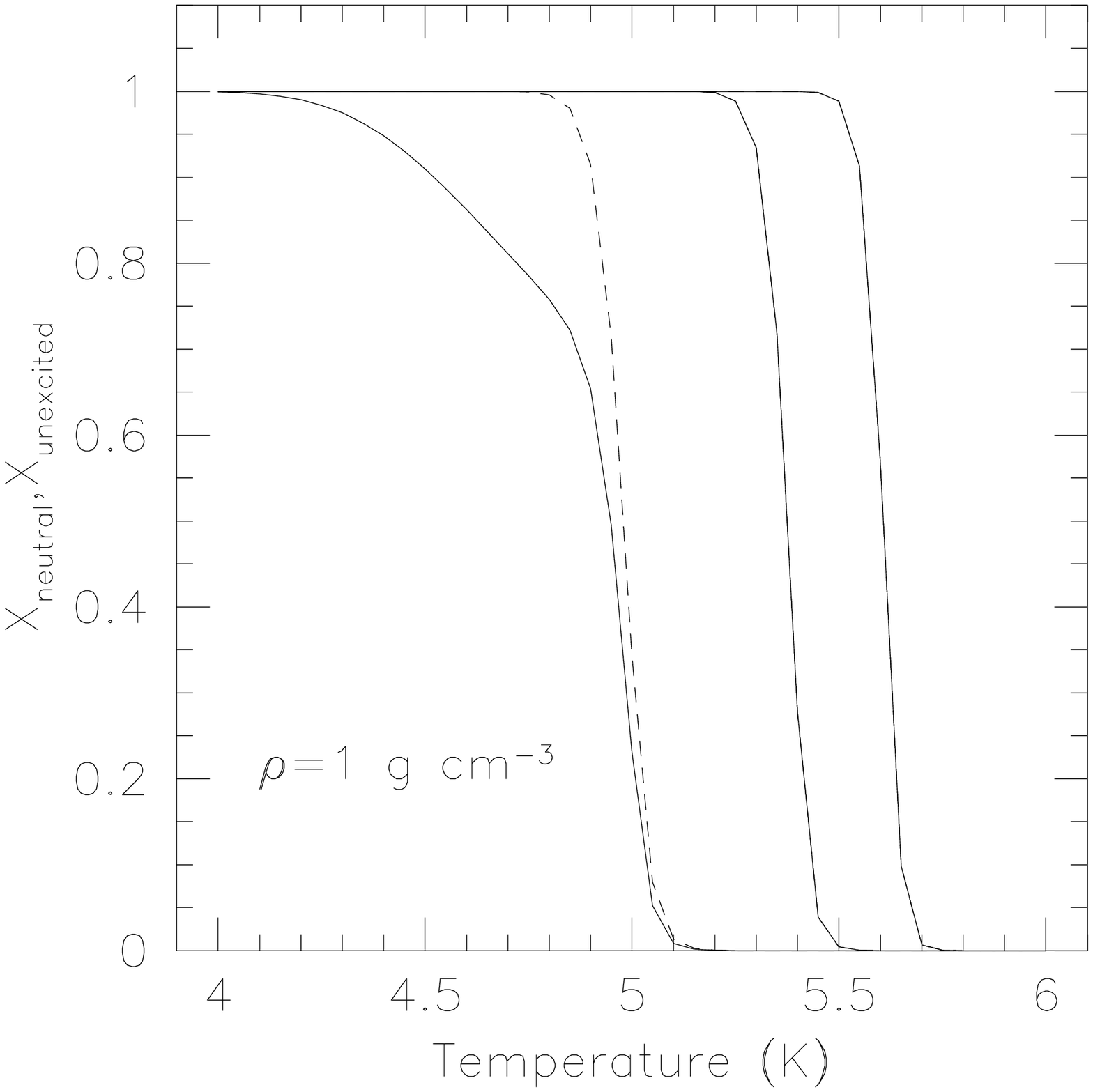}{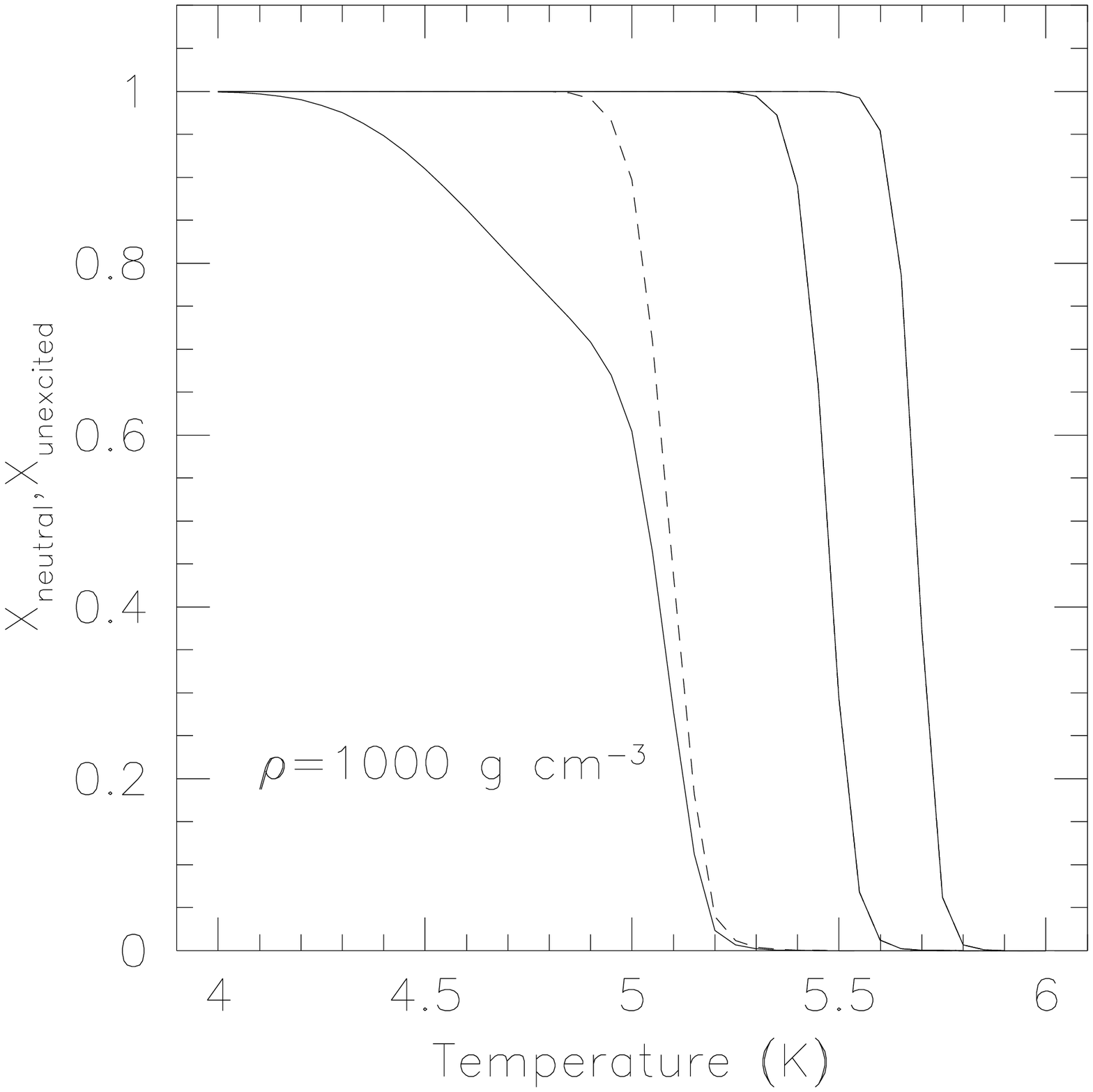}
\caption{The ground-state fraction as a function of temperature, density and
magnetic field.  The left panel shows the neutral fraction as a dashed
line and the unexcited fraction as a solid line for $\rho \sim
1$~g~cm$^{-3}$ and $B=10^{12}, 10^{14}$ and $10^{16}$~G.  The right
panel is for a density $\rho \sim 1000$~g~cm$^{-3}$.}
\label{fig:neufrac}
\end{figure}

Combining \eqref{groundeq}~and~\eqref{ioneq} yields the
fraction of ``shielded'' nuclei as a a function of temperature,
density and magnetic-field strength.  \figref{neufrac} depicts
the fraction of unexcited hydrogen atoms in the gas as function of
temperature for several field strengths and two densities.

\subsection{Thermonuclear reactions}

\begin{figure}
\plottwo{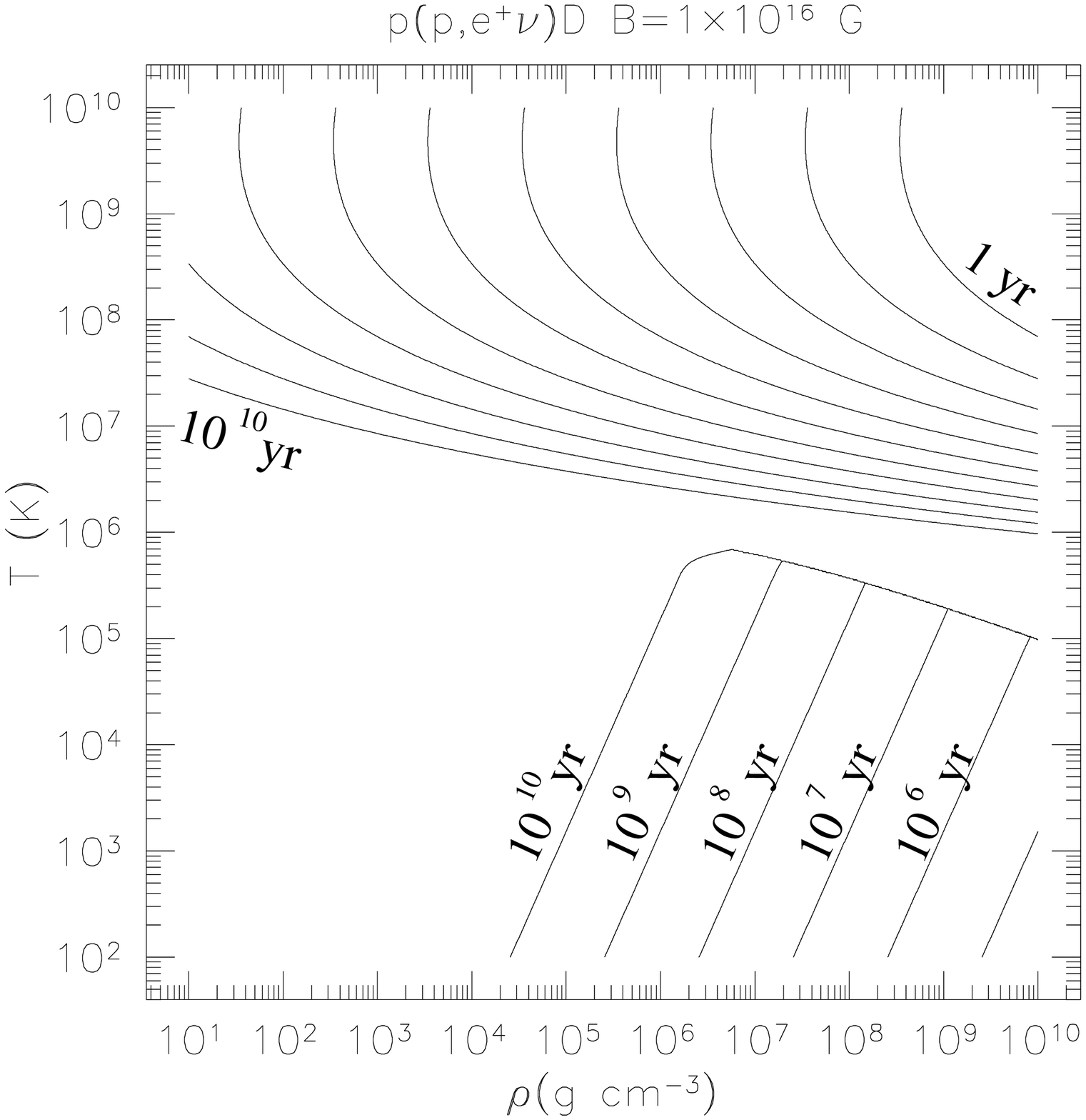}{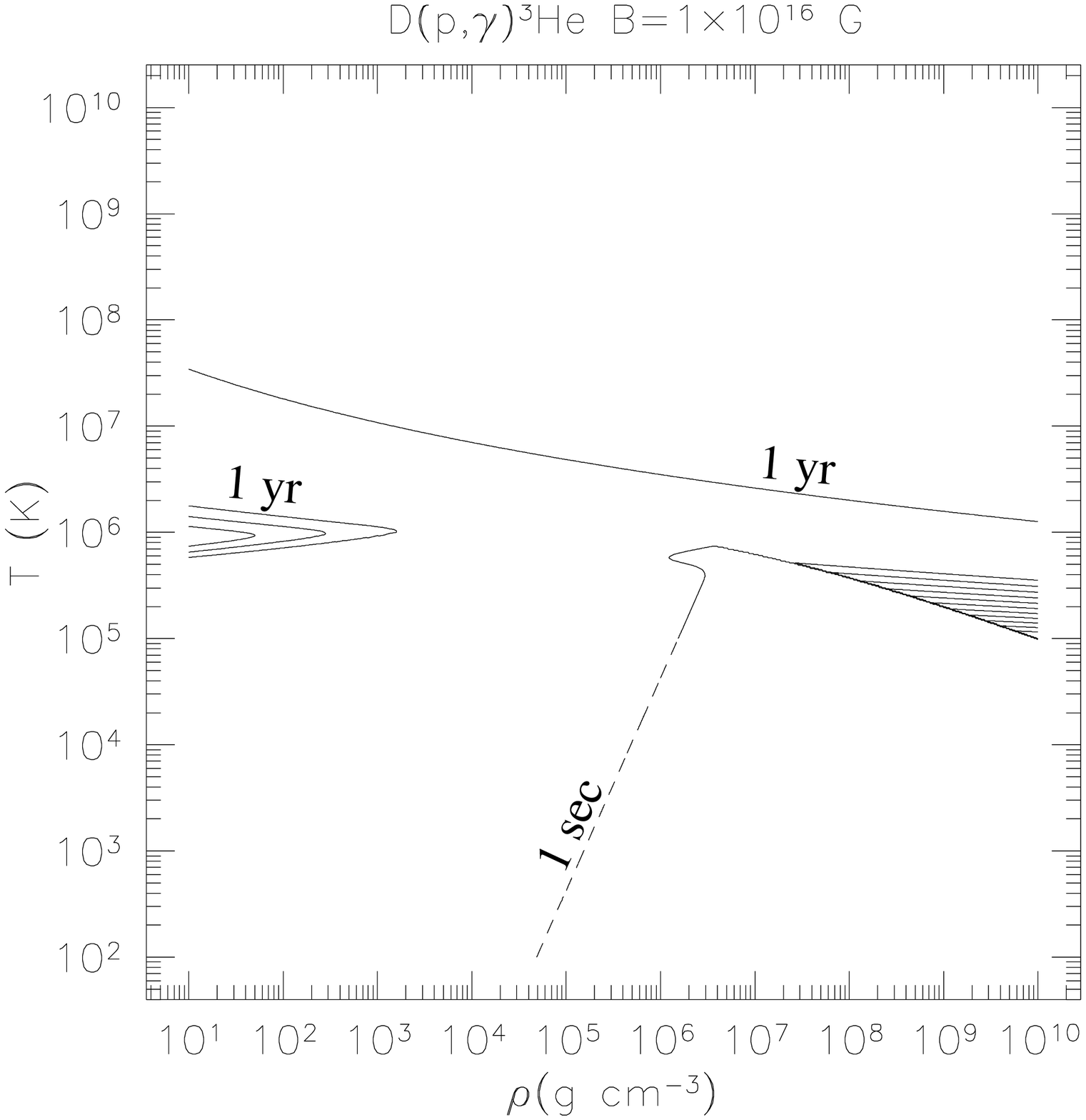}
\caption{The two panels depict the reaction timescale for the reactions 
$p(p,e^+\nu)D$ and $D(p,\gamma)^3$He for $B=10^{16}$G over a range of 
temperatures and densities.  The dashed contour traces $\tau$ of one second.  
The solid contours trace locii of timescales ranging from one year to 
$10^{10}$ years with a factor of ten in between each contour.}
\label{fig:tau16}
\end{figure}

We parameterize the thermonuclear reaction rates (\eg \cite{Clay83}) by
\begin{eqnarray}
r_{pp} & = & 3.06 \times 10^{-37} \rmmat{cm}^3 \rmmat{sec}^{-1}
	n_p^2 T_6^{-2/3} \exp(-33.71 T_6^{-1/3}) \\
r_{pD} & = & 3.28 \times 10^{-19} \rmmat{cm}^3 \rmmat{sec}^{-1}
	n_p n_D T_6^{-2/3} \exp(-37.11 T_6^{-1/3}).
\end{eqnarray}
The timescale for the exhaustion of a particular reactant becomes
\be
\tau_1 = { n_1 \over r_\rmscr{thermo} + r_\rmscr{magneto} }.
\ee
\figref{tau16} shows the reaction timescale for the consumption of 
hydrogen and deuterium in the reactions $p(p,e^+\nu)D$ and $D(p,\gamma)^3$He 
respectively for a magnetic field of $10^{16}$G.

Even in this very strong magnetic field, the p-p reaction proceeds only very 
slowly below temperatures of one million degrees; however, over millions of 
years, the hydrogen gas would be processed to deuterium and then to helium 
in such a strong magnetic field.  It would provide a steady source of energy, 
while eroding the storehouse of hydrogen which could potentially fuel a 
thermonuclear runaway.  Relatively, the second reaction proceeds instantly 
with timescales of less than one year for the interesting range of densities
and temperatures.  

\begin{figure}
\plottwo{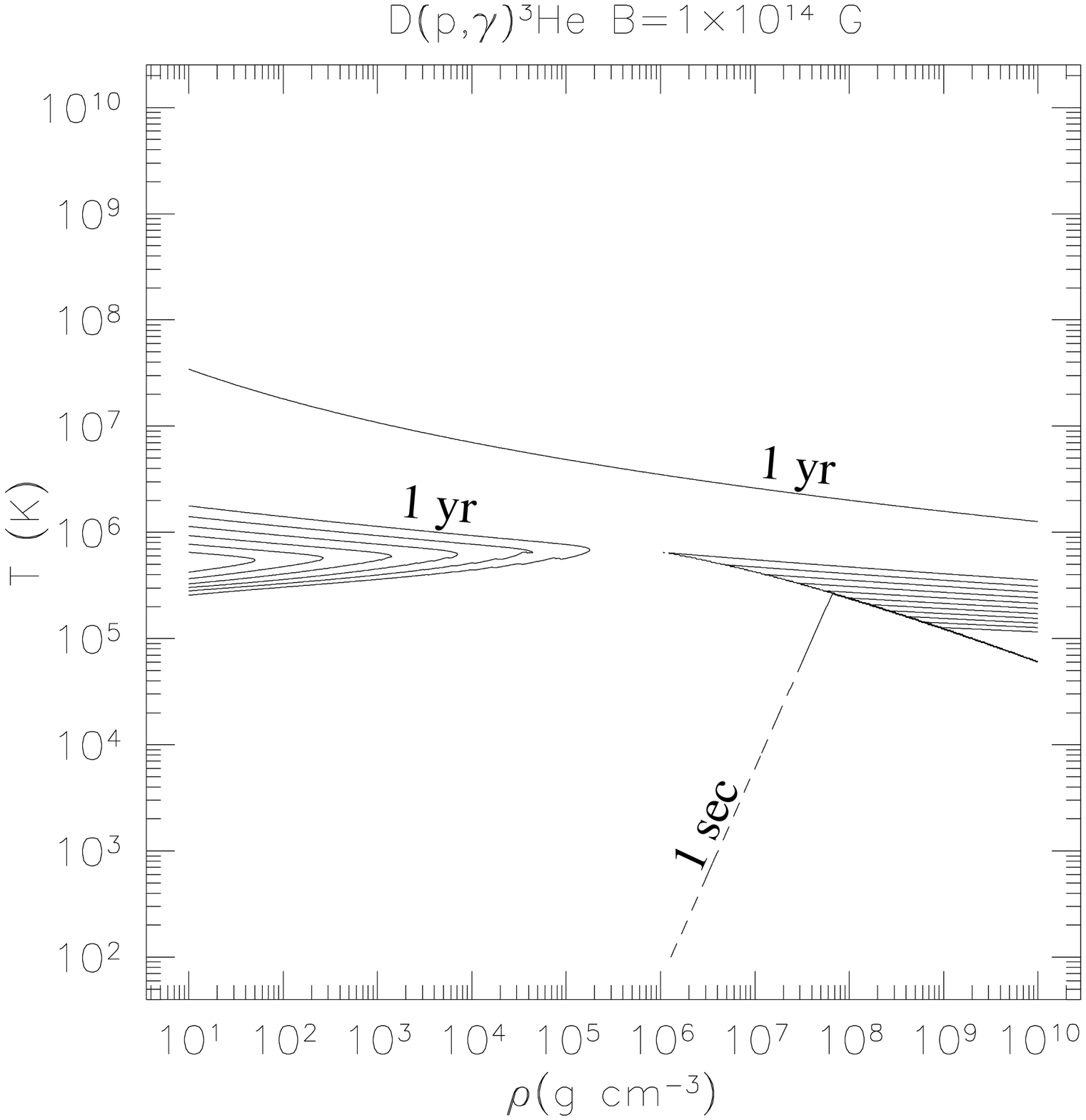}{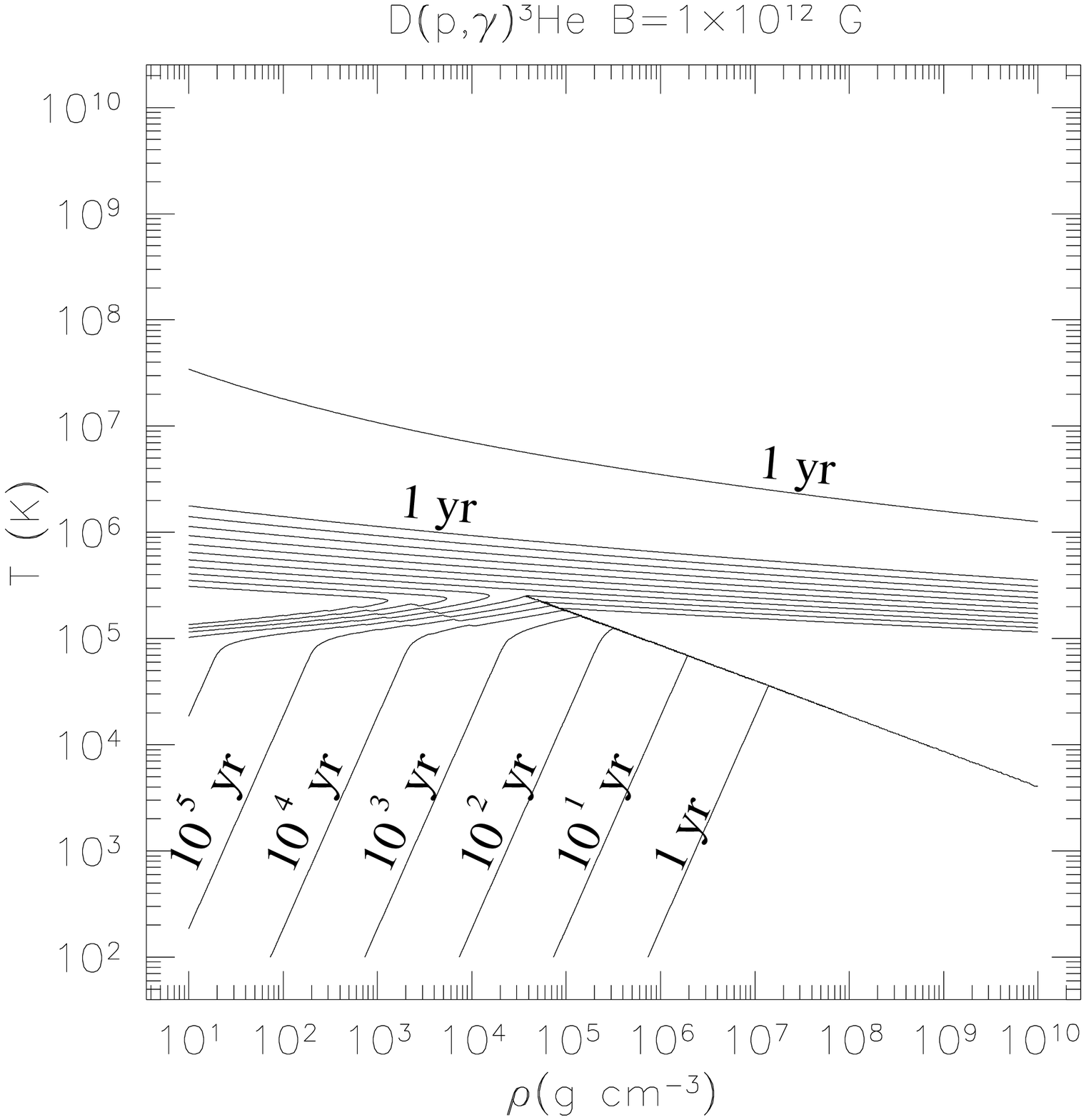}
\caption{The two panes depict the reaction timescale for the reaction
$D(p,\gamma)^3He$ for $B=10^{14}$G and $10^{12}$G over a range of 
temperatures and densities.  The dashed contour traces $\tau$ of one second.  
The solid contours trace locii of timescales ranging from one year to 
$10^{10}$ years with a factor of ten in between each contour.}
\label{fig:taud}
\end{figure}

For the weak fields depicted in \figref{taud} only the deuterium 
reaction proceeds at a significant rate.

\section{Discussion}

We find that in strong magnetic fields ($B \gsim 10^{12}$ G), the
cross-section for nuclear fusion is dramatically larger than in the
unmagnetized case.  For these strong fields, deuterons
fuse to $^3$He over short timescales ($\lsim 10^6$ yr) for the
density and temperatures expected on the surface of a neutron star.
Because of the inherent weakness of the $p-p$ interaction, the fusion
of protons to deuterium is only important over cosmological timescales
for ultrastrong fields ($B \gsim 10^{16}$ G) in spite of the large
enhancement in the cross section of this reaction.

For larger atoms ($Z>1$), we expect that reaction cross-sections will
also be larger in the presence of an intense magnetic field.
However, the shielding is unlikely to be as effective as for the $Z=1$
case, because additional electrons must occupy $m>0$ levels which are
much less effective at screening the nuclear charge.

\acknowledgements

This material is based upon work supported under a National Science
Foundation Graduate Fellowship.

\end{document}